\documentclass[10pt]{article}
\usepackage{jheppub}
\makeatletter
\def\@fpheader{\relax}
\def\@fpheader{~}
\makeatother
\usepackage{bm}
\usepackage{dsfont}
\usepackage{mathrsfs}
\usepackage{empheq}
\usepackage{amsmath,amsfonts}
\usepackage{verbatim}
\usepackage{bibentry}
\usepackage{slashed}
\usepackage{multirow}
\usepackage{tikz}
\usepackage{array}
\allowdisplaybreaks
\begin{document}
\title{Interpolating Boundary Conditions on $AdS_2$}
\author[a]{Anthonny F. Canazas Garay,}
\author[b]{Diego H. Correa,}
\author[a]{Alberto Faraggi,}
\author[b]{Guillermo A. Silva}
\affiliation[a]{Departamento de Ciencias F\'isicas, Facultad de Ciencias Exactas, Universidad Andr\'es Bello,  \\ Sazi\'e 2212, Piso 7, Santiago, Chile.}
\affiliation[b]{Instituto de F\'isica La Plata - CONICET \& \\ Departamento de F\'isica, Universidad Nacional de La Plata, \\ C.C. 67, 1900, La Plata, Argentina}
\emailAdd{a.canazasgaray@uandresbello.edu}
\emailAdd{correa@fisica.unlp.edu.ar}
\emailAdd{alberto.faraggi@unab.cl}
\emailAdd{silva@fisica.unlp.edu.ar}

\abstract{We consider two instances of boundary conditions for massless scalars on $AdS_2$ that interpolate between the Dirichlet and Neumann cases while preserving scale invariance. Assessing invariance under the full $SL(2;\mathds{R})$ conformal group is not immediate given their non-local nature. To further clarify this issue, we compute holographically 2- and 4-point correlation functions using the aforementioned boundary conditions and study their transformation properties. Concretely, motivated by the dual description of some multi-parametric families of Wilson loops in ABJM theory, we look at the excitations of an open string around an $AdS_2\subset AdS_4\times\mathbb{CP}^3$ worldsheet, thus obtaining correlators of operators inserted along a $1$-dimensional defect in ${\cal N}=6$ super Chern-Simons-matter theory at strong coupling. Of the two types of boundary conditions analyzed, only one leads to the expected functional structure for conformal primaries; the other exhibits covariance under translations and rescalings but not under special conformal transformations.
}
\maketitle
\section{Introduction}\label{Sec: Intro}

The existence of multi-parametric families of supersymmetric Wilson loops is a remarkable feature of ABJM theories \cite{Ouyang:2015bmy,Ouyang:2015iza} (see \cite{Drukker_2020} for a review of ABJM Wilson loops). In particular, there exists a one-parameter family that interpolates between the $1/2$-BPS {\it fermionic} \cite{Drukker_2008,CHEN201038,Rey_2009} and the $1/6$-BPS {\it bosonic} Wilson loops \cite{Drukker_2010}. The former were put in correspondence with open strings satisfying Dirichlet boundary conditions in all the angular coordinates, while the latter correspond to imposing Neumann conditions on two directions along a $\mathbb{CP}^1\subset\mathbb{CP}^3$. Given these maps, it is quite natural to associate the entire parametric family of Wilson loops with open strings in $AdS_4\times \mathbb{CP}^3$ satisfying some kind of interpolating boundary conditions that connect the Dirichlet and Neumann cases. Although a proposal was put forward in \cite{Correa_2020}, the precise form of these boundary conditions is not completely understood. One question that remains unanswered is whether the Wilson loops preserve the full conformal group, and not just scale invariance, for any value of the interpolating parameter. Thus, discovering instances of interpolating boundary conditions on $AdS_2$ compatible with $SL(2;\mathds{R})$ conformal symmetry might shed some light on the identification of the holographic dual for this family of Wilson loops. This is one of the main motivations for the present work.

The guiding principle for identifying the interpolating boundary conditions is that they must preserve both supersymmetry and scale invariance. With regard to supersymmetry, it was shown in \cite{Correa_2020} that the existence of a rich moduli of supersymmetric boundary conditions can be associated to the appearance of massless fermions in the spectrum of IIA strings in $AdS_4\times \mathbb{CP}^3$ (see \cite{Correa:2021sky} for the same phenomenon in a different setup). This is contrary to the case of $AdS_5\times S^5$ (cf. \cite{Polchinski:2011im,Faraggi:2011bb,Faraggi:2016ekd}) where all fermionic fluctuations dual to the $1/2$-BPS Wilson loop in $\mathcal{N}=4$ SYM are massive. On the other hand, the reason to search for scale-preserving boundary conditions is that it is expected that each Wilson loop in the family corresponds to a defect CFT$_1$. This arises from the fact that the Wilson loop vevs are independent of the interpolation parameter $\zeta$ \cite{OUYANG2016215,OUYANG2016496}. As argued in \cite{Klebanov:2011gs,Fei:2015oha,Beccaria:2017rbe} for the case of  circular loops, the beta function for $\zeta$ is proportional to the derivative of the CFT$_1$ free energy (given by the log of the Wilson loop vev) with respect to $\zeta$. The independence of the vev on the interpolating parameter can then be interpreted as a sign of scale invariance. 

Specifying interpolating boundary conditions that preserve conformal symmetry is a subtle problem. As is well known, changes in the boundary conditions for fields in AdS entail deformations of the dual CFT which usually break scale invariance. As an example consider a massive scalar on $AdS_{d+1}$ \cite{Malda,GKP,Witten:1998qj}, whose asymptotic behavior is
\begin{empheq}{alignat=7}\label{scalarfieldexpansion}
	\phi(z,x)&\underset{z\to0}{\longrightarrow}z^{\Delta_-}\left(\alpha(x)+\cdots\right)+z^{\Delta_+}\left(\beta(x)+\cdots\right)\,,
	&\qquad
	\Delta_{\pm}&=\frac{d}{2}\pm\sqrt{\frac{d^2}{4}+m^2R^2}\,.
\end{empheq}
For $-\frac{d^2}{4}<m^2R^2<-\frac{d^2}{4}+1$, one can impose either Dirichlet or Neumann boundary conditions, i.e. fix $J(x)=\alpha(x)$ or $J(x)=\beta(x)$, giving rise to CFT operators with scale dimensions $\Delta_+$ or $\Delta_-$, respectively. This range of masses is known as the Breitenlohner-Freedman (BF) window \cite{BREITENLOHNER1982197,BREITENLOHNER1982249}. Arbitrary combinations of  $\alpha(x)$ and  $\beta(x)$ which interpolate between Dirichlet and Neumann boundary conditions are also permitted, and they can be seen to correspond to multi-trace deformations of the dual CFT.\footnote{Supersymmetric multi-trace boundary conditions for scalar supermultiplets on $AdS_{d+1}$ were considered in \cite{Amsel_2009} for dimensions $d=2\,,3\,,4$.} For instance, setting $J(x)=\alpha(x)+\chi\beta(x)$ as the source describes an interpolation between two fixed points. Since $\alpha(x)$ and $\beta(x)$ have different mass dimensions, this choice introduces a dimensionful parameter $\chi$ into the problem.  As a result, scale invariance is broken by the boundary condition and a renormalization group flow in the dual field theory is triggered \cite{Witten:2001ua,Hartman:2006dy,GM}.

The case of interest to us is that of massless scalar fields in $AdS_2$, as these account for the angular fluctuations of the string worldsheet dual to the supersymmetric Wilson loops.  Imposing either Neumann or Dirichlet boundary conditions on them should correspond, in the dual description, to primary operators in one dimension respecting conformal symmetry. Indeed, correlation  functions for operator insertions using Dirichlet boundary conditions  have been recently computed in \cite{Bianchi_2020}, finding full agreement with field theory expectations.

The boundary conditions proposed in \cite{Correa_2020} compensate for the difference in scale dimensions between $\alpha(\tau)$ and $\beta(\tau)$ by taking the derivative of $\alpha(\tau)$ with respect to the boundary coordinate $\tau$, rendering the interpolating parameter dimensionless. An integrated version of that boundary condition reads
\begin{empheq}{alignat=7}
	\label{eq: definition J free original}
J(\tau)&\equiv\cos\chi\,\alpha(\tau)+\frac{1}{2}\sin\chi\int_{-\infty}^{\infty}d\tau'\beta(\tau'){\rm sign}(\tau-\tau')\,.
\end{empheq}
However, having a dimensionless parameter is not sufficient to guarantee full conformal invariance at the quantum level. One way of testing if a set of boundary conditions is compatible with a given symmetry is by computing correlation functions holographically. As it turns out, the 2-point functions that result from sourcing the combination \eqref{eq: definition J free original} are consistent with conformal symmetry. To further understand if this invariance is actually present in the dual defect theory it is important to study higher-point correlation functions, as these could lead to more stringent tests. Our goal is then to compute holographic $4$-point functions associated to interpolating boundary conditions of the type \eqref{eq: definition J free original} and check whether they respect the conformal structure or not. In particular, for a CFT defined on an infinite line, a primary operator ${\cal O}(\tau)$ of scale dimension $\Delta$ has a $4$-point correlator of the form
\begin{empheq}{alignat=7}\label{4ptform}
    \langle{\cal O}(\tau_1){\cal O}(\tau_2){\cal O}(\tau_3){\cal O}(\tau_4)\rangle&=\frac{G(u)}{(\tau_1-\tau_2)^{2\Delta}(\tau_3-\tau_4)^{2\Delta}}\,,
    &\qquad
    u&=\frac{(\tau_1-\tau_2)(\tau_3-\tau_4)}{(\tau_1-\tau_3)(\tau_2-\tau_4)}\,,
\end{empheq}
where $G(u)$ is an arbitrary function of the unique independent cross-ratio, $u$, that exits in 1d. Full conformal symmetry and not just scale invariance is needed to conclude this. 

As we will see, our main result in this paper is that the holographic 4-point function resulting from the boundary condition \eqref{eq: definition J free original} does not respect the form \eqref{4ptform}. Although this is not a priori an impediment for its interpretation as the dual to the family of Wilson loops under scrutiny, it is of interest to consider alternatives that are actually consistent with full conformal symmetry. Thus, we will allow for the possibility of a second kind of boundary condition by replacing the sign function in \eqref{eq: definition J free original} with another dimensionless function.

The rest of paper is organized as follows. In section \ref{VPReview} we review the implementation of boundary conditions through the addition of boundary terms to the action consistent with the variational principle. We do this for the source \eqref{eq: definition J free original} as well as for an alternative interpolating boundary condition, also consistent with scale invariance. Section \ref{sec: correlation functions} is devoted to the computation of Witten digrams in $AdS_2$ using these interpolating boundary conditions, thus obtaining correlation functions of excitations in a 1d defect in the strong coupling limit. We first review the computation of the 2-point function and then turn to the calculation of the 4-point function using the quartic interactions derived from the Nambu-Goto action. In section \ref{sec: discussion}, we conclude discussing our results.

\section{Interpolating Boundary Conditions}
\label{VPReview}
We  work in Euclidean $AdS_2$ space with unit radius. In Poincar\'e coordinates $(z,\tau)$ the metric takes the form
\begin{empheq}{alignat=7}\label{eq: AdS2 metric}
	ds^2&=\frac{1}{z^2}\left(d\tau^2+dz^2\right)\,,
	&\qquad
	z&>0\,,
	&\quad
	-\infty&<\tau<\infty\,.
\end{empheq}
To regulate possible divergences the boundary is located at $z=\epsilon$, the induced metric is $h=\epsilon^{-2}$ and the outer normal vector becomes $\partial_n=-z\partial_z$.

\subsection{Variational Principle and AdS/CFT}\label{subsec: variational}

Let us start by reviewing the role that boundary terms and the variational principle play in the calculation of correlation functions in AdS/CFT \cite{min1,min2,KW,SK02}. Consider a massless complex scalar field in $AdS_2$ with action 
\begin{empheq}{alignat=7}\label{eq: S0}
	S_0&=\int d^2x\sqrt{g}\,\partial_{\mu}\bar{\phi}\partial^{\mu}\phi\,.
\end{empheq}
Setting $d=1$ and $m=0$ in \eqref{scalarfieldexpansion} we find that the possible conformal dimensions are $\Delta_-=0$ and $\Delta_+=1$. The asymptotic expansion then reads
\begin{empheq}{alignat=7}
\phi(z,\tau)&=\alpha(\tau)+z\beta(\tau)+\mathcal{O}(z^2)\,.
\end{empheq}
Assuming regularity in the bulk  ($z\to\infty$), the on-shell variation of the action takes the form
\begin{empheq}{alignat=7}\label{eq: variation S0}
	\delta S_0&=\int_{-\infty}^{\infty}d\tau\sqrt{h}\left(\partial_n\bar{\phi}\delta\phi+\partial_n\phi\delta\bar{\phi}\right)\Big|_{z=\epsilon}&&=-\int_{-\infty}^{\infty}d\tau\left(\bar{\beta}(\tau)\delta\alpha(\tau)+\beta(\tau)\delta\bar{\alpha}(\tau)\right)\,.
\end{empheq}
For massless fields we can safely take the regulating parameter $\epsilon\to0$ since no divergences arise. We learn from \eqref{eq: variation S0} that   $S_0$ is appropriate for a variational problem in which the function $\alpha(\tau)$ is fixed; only then does \eqref{eq: S0} have an actual extremum when $\Box\phi=0$. This corresponds to the usual Dirichlet boundary conditions in AdS. Moreover, according to the AdS/CFT dictionary, the $1$-point function (or vev) of the dual operator ${\cal O}_D$ in the presence of the source $J_D \equiv\alpha $ is
\begin{empheq}{alignat=7}\label{eq: vev Dirichlet}
	\langle\bar{\mathcal{O}}_D(\tau)\rangle&\equiv-\frac{\delta S_D[J]}{\delta J_D(\tau)}&&=\bar{\beta}(\tau)\,,
    &\qquad
    S_D&\equiv S_0\,.
\end{empheq}
Alternatively, Neumann boundary conditions require the addition of the boundary term
\begin{empheq}{alignat=7}\label{eq: Sbdry Neumann}
	S^N_{\textrm{bdry}}&=-\int_{-\infty}^{\infty}d\tau\sqrt{h}\left(\phi\partial_n\bar{\phi}+\bar{\phi}\partial_n\phi\right)\Big|_{z=\epsilon}&&=\int_{-\infty}^{\infty}d\tau\left(\alpha(\tau)\bar{\beta}(\tau)+\bar{\alpha}(\tau)\beta(\tau)\right)\,.
\end{empheq}
Evaluating the action on-shell one finds
\begin{empheq}{alignat=7}
	\delta\left(S_0+S^N_{\textrm{bdry}}\right)&=\int_{-\infty}^{\infty}d\tau\left(\bar{\alpha}(\tau)\delta\beta(\tau)+\alpha(\tau)\delta\bar{\beta}(\tau)\right)\,.
\end{empheq}
Hence, the $1$-point function of the dual operator ${\cal O}_N$ sourced by $J_N \equiv\beta $ is
\begin{empheq}{alignat=7}\label{eq: vev Neumann}
	\langle\bar{\mathcal{O}}_N(\tau)\rangle&\equiv-\frac{\delta S_N[J]}{\delta J_N(\tau)}&&=-\bar{\alpha}(\tau)\,,
    &\qquad
    S_N&\equiv S_0+S^N_{\textrm{bdry}}\,.
\end{empheq}
The possibility of imposing either boundary condition is due to the fact that a massless scalar field in $AdS_2$ lies inside the BF window.

\subsection{Local and Non-local Boundary Conditions}\label{subsec: bc}

Generically, Dirichlet and Neumann boundary conditions are compatible with the isometries of $AdS_2$, leading to correlation functions that exhibit 1d conformal symmetry.  Moreover, when appropriately combined with fermionic fields, they also respect supersymmetry \cite{Sakai:1984vm}. In this paper we   explore boundary conditions that interpolate between the Dirichlet and Neumann cases while preserving scale invariance, and wonder whether they are consistent with the full conformal group.

To achieve our goal we need a well-posed variational problem in which the boundary source $J(\tau)$ combines the two fall-off functions $\alpha(\tau)$ and $\beta(\tau)$ in a suitable way. Since  these fields have different  mass  dimensions, directly adding them necessarily breaks scale invariance \cite{Witten:2001ua,Hartman_2008}. A specific $AdS_2$ interpolation of this kind was considered in \cite{Polchinski:2011im} and interpreted as a Renormalization Group flow in the dual CFT$_1$. As shown in \cite{Correa_2020}, however, the  difference in scale  dimensions can be compensated by  combining  $\beta(\tau)$ with the derivative of $\alpha(\tau)$ or, more conveniently (for reasons that will become clear below), $\alpha(\tau)$ with an integral of $\beta(\tau)$. Following these insights, we propose boundary conditions defined by fixing the combination\footnote{In the following, the Cauchy principal value is implicit in all integrals in order to deal with divergences at $\tau=\tau'$.}
\begin{empheq}{alignat=7}
	\label{eq: definition J free}
J(\tau)&\equiv\cos\chi\,\alpha(\tau)+\sin\chi\int_{-\infty}^{\infty}d\tau'\beta(\tau')g(\tau-\tau')\,,
\end{empheq}
where $\chi\in \left[0,\frac{\pi}{2}\right]$ is an interpolating parameter and $g(\tau)$ is a real dimensionless function we will promptly identify. Equation \eqref{eq: definition J free} includes \eqref{eq: definition J free original} as a particular case; it also enables us to define alternative interpolating boundary conditions. The main requisite is that $g(\tau)$ satisfy the closure relation
\begin{empheq}{alignat=7}\label{eq: closure}
	\int_{-\infty}^{\infty}d\tau''\partial_{\tau}g(\tau-\tau'')\partial_{\tau'}g(\tau'-\tau'')=\delta(\tau-\tau')\,,
\end{empheq}
which allows us to implement the boundary condition in the variational problem via\footnote{Recall that in Euclidean signature hermitian conjugation acts as $\bar{f}(\tau)\equiv f(-\tau)^*$. It is easy to see that \eqref{eq: Sbdry free} is real.}
\begin{empheq}{alignat=7}\label{eq: Sbdry free}
	S_{\textrm{bdry}}&\equiv-\cos\chi\sin\chi\int_{-\infty}^{\infty}d\tau\int_{-\infty}^{\infty}d\tau'\left(\bar{\beta}(\tau)\beta(\tau')-\partial_{\tau'}\bar{\alpha}(\tau')\partial_{\tau}\alpha(\tau)\right)g(\tau-\tau')
	\cr
	&\hspace{11pt}+\sin^2\chi\int_{-\infty}^{\infty}d\tau\left(\bar{\alpha}(\tau)\beta(\tau)+\alpha(\tau)\bar{\beta}(\tau)\right)\,.
\end{empheq}
Alternatively, this can be written as
\begin{empheq}{alignat=7}\label{eq: Sbdry free contact}
	S_{\textrm{bdry}}&=-\tan\chi\int_{-\infty}^{\infty}d\tau\int_{-\infty}^{\infty}d\tau'\left(\bar{\beta}(\tau)\beta(\tau')-\partial_{\tau'}\bar{J}(\tau')\partial_{\tau}J(\tau)\right)g(\tau-\tau')\,.
\end{empheq}
Neglecting a total $\tau$-derivative the  variation of the on-shell action then reads
\begin{empheq}{alignat=7}\label{eq: variation S free}
	\delta\left(S_0+S_{\textrm{bdry}}\right)&=-\int_{-\infty}^{\infty}d\tau\left(\langle\bar{\mathcal{O}}(\tau)\rangle\delta J(\tau)+\langle\mathcal{O}(\tau)\rangle\delta\bar{J}(\tau)\right)\,,
\end{empheq}
where
\begin{empheq}{alignat=7}
	\label{eq: definition vev free}
	\langle\bar{\mathcal{O}}(\tau)\rangle&\equiv\cos\chi\,\bar{\beta}(\tau)+\sin\chi\int_{-\infty}^{\infty}d\tau'\partial_{\tau'}\bar{\alpha}(\tau')\partial_{\tau}g(\tau-\tau')\,.
\end{empheq}
We see that $S_0+S_{\textrm{bdry}}$ is the  appropriate action for the variational problem at hand, namely, that in which the combination \eqref{eq: definition J free} is fixed. Moreover, according to the AdS/CFT dictionary, the $1$-point function of the dual operator sourced by $J(\tau)$ is given by expression \eqref{eq: definition vev free}. As $\chi$ varies, the source and vev interpolate between the Dirichlet case and a $g$-transformed version of the Neumann case (see discussion at the end of this section). We remark that in the AdS/CFT correspondence boundary actions are defined only up to arbitrary functionals of the sources. In writing \eqref{eq: Sbdry free} and \eqref{eq: Sbdry free contact} this ambiguity was fixed by demanding that not only $J(\tau)$ has the correct Dirichlet and Neumann limits, but also $\langle\bar{\mathcal{O}}(\tau)\rangle$ does. As a consistency check, notice that $S_{\textrm{bdry}}$ vanishes for $\chi=0$ and reduces to $S^N_{\textrm{bdry}}$ for $\chi=\frac{\pi}{2}$.

Some comments regarding the function $g(\tau)$ are now in order. We consider two alternatives, dubbed \emph{local} (L) and \emph{non-local} (NL):
\begin{empheq}{alignat=7}\label{eq: definition of g}
\begin{split}
	g(\tau)&=\left\{
	\begin{array}{>{\displaystyle}c>{\displaystyle}l}
		\frac{1}{2}\textrm{sign}(\tau)\,, & \quad\textrm{L}
		\\\\
		\frac{1}{\pi}\ln\left(|\tau|\right)\,, & \quad\textrm{NL}
	\end{array}
	\right.
\end{split}
	\qquad\Rightarrow\qquad
\begin{split}
	\partial_{\tau}g(\tau)&=\left\{
	\begin{array}{>{\displaystyle}c>{\displaystyle}l}
		\delta(\tau)\,, & \quad\textrm{L}
		\\\\
	\frac{1}{\pi\tau}\,, & \quad\textrm{NL}
	\end{array}
	\right.\,.
\end{split}
\end{empheq}
This terminology stems from the fact that upon taking a derivative of \eqref{eq: definition J free} we obtain
\begin{empheq}{alignat=7}\label{eq: diff J free}
	\partial_{\tau}J(\tau)&=\left\{
	\begin{array}{>{\displaystyle}c>{\displaystyle}l}
		c_{\chi}\partial_{\tau}\alpha(\tau)+s_{\chi}\beta(\tau)\,, & \quad\textrm{L}
		\\\\
		c_{\chi}\partial_{\tau}\alpha(\tau)+s_{\chi}\hat{\beta}(\tau)\,, & \quad\textrm{NL}
	\end{array}
	\right.\,,
	&\qquad
	\langle\bar{\mathcal{O}}(\tau)\rangle&=\left\{
	\begin{array}{>{\displaystyle}c>{\displaystyle}l}
		c_{\chi}\bar{\beta}(\tau)+s_{\chi}\partial_{\tau}\bar{\alpha}(\tau)\,, & \quad\textrm{L}
		\\\\
		c_{\chi}\bar{\beta}(\tau)+s_{\chi}\partial_{\tau}\hat{\bar{\alpha}}(\tau)\,, & \quad\textrm{NL}
	\end{array}
	\right.\,.
\end{empheq}
Here we have abbreviated $c_\chi=\cos\chi$ and $s_\chi=\sin\chi$, and denoted by $\hat{\beta}(\tau)$ the Hilbert transform of $ {\beta}(\tau)$, defined as (see appendix \ref{app: Hilbert Transform})
\begin{empheq}{alignat=7}
    \hat{\beta}(\tau)&\equiv\frac{1}{\pi}\textrm{p.v.}\int_{-\infty}^{\infty}dt\frac{\beta(t)}{\tau-t}\,.
\end{empheq}
So, even though the current in \eqref{eq: definition J free} sources a non-local combination of $\alpha(\tau)$ and $\beta(\tau)$, its derivative is  local in these fields for the L choice of boundary conditions. The same is true for the vev in \eqref{eq: diff J free}. 
As mentioned above, the L boundary condition is equivalent to one of the cases considered in \cite{Correa_2020}. 
The NL alternative is a new (non-local) boundary condition for massless fields on $AdS_2$ that, as we will see, preserves the conformal invariance in the dual CFT$_1$.

The Hilbert transform is a mathematical tool widely used in signal processing and other areas of physics \cite{King_2009}, and its appearance in the present setting is quite natural. Recall that the fall-off fields $\alpha(\tau)$ and $\beta(\tau)$ become linked by  regularity of $\phi(z,\tau)$ in the interior of AdS ($z\to\infty$). Indeed, the most general regular solution to the equation of motion $\Box\phi=0$ can be written as 
\begin{empheq}{alignat=7}
	\phi(z,\tau)&=\frac{1}{\sqrt{2\pi}}\int_{-\infty}^{\infty}dw\,e^{-|w|z+iw\tau}\tilde\phi(w)\,,
\end{empheq}
with $\tilde\phi(w)$ an arbitrary function. Using the fact that the Hilbert transform takes $e^{iw\tau}\mapsto -i\,\textrm{sign}(w)e^{iw\tau}$, it is straightforward to see that
\begin{empheq}{alignat=7}
	\partial_z \hat\phi(z,\tau)&=\partial_\tau \phi(z,\tau)
	&\qquad\Rightarrow\qquad   
	\hat\beta(\tau)&=\partial_{\tau}{\alpha}(\tau)\,.
	\label{betahatalpha}
\end{empheq}
Since $\hat\beta(\tau)$ and $\partial_{\tau}{\alpha}(\tau)$ are locally related for regular solutions, it seems reasonable to combine them as in \eqref{eq: diff J free}.

\section{Correlation Functions at strong coupling}\label{sec: correlation functions}
We now move on to study the tree-level correlation functions that result from applying the AdS/CFT dictionary to the interpolating boundary conditions introduced in the previous section. To this purpose we consider the dynamics arising from the fluctuations of a type IIA open string around an $AdS_2\subset AdS_4\times\mathds{CP}^3$ classical worldsheet. The bosonic spectrum includes two transverse fluctuations in $AdS_4$ with $m^2=2$ and six fluctuations along $\mathbb{CP}^3$ with $m^2=0$. We identify the scalar field discussed in previous section with a complex combination of these massless excitations, which can be put in correspondence with certain components of the displacement (super) multiplet along the Wilson loop. The expansion of the Nambu-Goto action to fourth order in the effective string tension was performed in \cite{Bianchi_2020}. Interestingly, the quartic interactions involve derivatives of the field. This forces us to revisit the variational problem and modify the definition of the source and vev. Performing a first principles derivation of the $4$-point function (relegated to appendix \ref{app: derivation}) we verify that the standard prescription in terms of Witten diagrams remains valid once the appropriate boundary terms are added to the action.

Before we proceed with the calculation of holographic correlation functions, let us explain how the Dirichlet and Neumann results can be recovered as limiting cases of our proposed interpolating boundary conditions. From \eqref{eq: definition J free} and \eqref{eq: definition vev free} we immediately see that
\begin{empheq}{alignat=7}
    J_D(\tau)&=J(\tau)\Big|_{\chi=0}\,,
    &\qquad
    \langle\bar{\mathcal{O}}_D(\tau)\rangle&=\langle\bar{\mathcal{O}}(\tau)\rangle\Big|_{\chi=0}\,,
\end{empheq}
so the $\chi=0$ limit will yield correlators corresponding to Dirichlet boundary conditions, for both the L and NL choices. On the other end of the interpolation, however, neither the source nor the vev directly reduce to their Neumann counterparts. Instead, the relation involves a derivative/Hilbert transform, namely,
\begin{empheq}{alignat=7}\label{eq: Neumann limit}
	J_N(\tau)&=\left\{
	\begin{array}{>{\displaystyle}r>{\displaystyle}l}
		\partial_{\tau}J(\tau)\Big|_{\chi=\frac{\pi}{2}} & \quad\textrm{L}
		\\\\
		-\partial_{\tau}\hat{J}(\tau)\Big|_{\chi=\frac{\pi}{2}} & \quad\textrm{NL}
	\end{array}
	\right.\,,
    &\qquad
    \partial_{\tau}\bar{\mathcal{O}}_N(\tau)&=\left\{
	\begin{array}{>{\displaystyle}r>{\displaystyle}l}
		-\bar{\mathcal{O}}(\tau)\Big|_{\chi=\frac{\pi}{2}} & \quad\textrm{L}
		\\\\
		\hat{\bar{\mathcal{O}}}(\tau)\Big|_{\chi=\frac{\pi}{2}} & \quad\textrm{NL}
	\end{array}
	\right.\,.
\end{empheq}
The connection between our proposal and the standard Neumann boundary conditions is found by considering $J'(\tau)\equiv\partial_{\tau}J(\tau)$ as the source when computing correlation functions. After all, we can always integrate by parts and write
\begin{empheq}{alignat=7}
    \delta S&\sim\int_{-\infty}^{\infty}d\tau\langle\bar{\mathcal{O}}(\tau)\rangle\delta J(\tau)&&=\int_{-\infty}^{\infty}d\tau\langle\bar{\mathcal{O}}'(\tau)\rangle\delta J'(\tau)\,,
\end{empheq}
with $\langle\bar{\mathcal{O}}(\tau)\rangle=-\langle\partial_{\tau}\bar{\mathcal{O}}'(\tau)\rangle$. The problem with the $J'(\tau)$-approach is that the dual operators $\bar{\mathcal{O}}'(\tau)$ have ``$\Delta_{\mathcal{O}'} =0$'', meaning they are not well-defined primaries (cf. \eqref{DvsN}). Using $J(\tau)$ as the source, on the other hand, yields correlators for the derivatives of such operators, $\mathcal{O}(\tau)=-\partial_{\tau}\bar{\mathcal{O}}'(\tau)$, which are proper primaries with $\Delta_{\mathcal{O}}=1$. In general, the relation between the two sets of correlation functions follows from noticing that, formally,\footnote{In Euclidean signature $\overline{\partial_{\tau}f(\tau)}=-\partial_{\tau}\bar{f}(\tau)$ and $\bar{\hat{f}}(\tau)=-\hat{\bar{f}}(\tau)$, hence the factors of $(-1)^n$.}
\begin{empheq}{alignat=7}
	\frac{\delta^{2n}S[J]}{\delta\bar{J}(\tau_1)\cdots\delta J(\tau_{2n})}&=(-1)^{n}\frac{\partial^{2n}}{\partial\tau_1\cdots\partial\tau_{2n}}\left(\frac{\delta^{2n}S'[J']}{\delta\bar{J}'(\tau_1)\cdots\delta J'(\tau_{2n})}\right)\,,
	\label{chain}
\end{empheq}
where $S'[J']\equiv S[J]$, up to total boundary derivatives. Therefore,
\begin{empheq}{alignat=7}
    \langle\mathcal{O}(\tau_1)\cdots\bar{\mathcal{O}}(\tau_{2n})\rangle&=(-1)^{n}\langle\partial_{\tau_1}\mathcal{O}'(\tau_1)\cdots\partial_{\tau_{2n}}\bar{\mathcal{O}}'(\tau_{2n})\rangle&&=\langle\partial_{\tau_1}\mathcal{O}'(\tau_1)\cdots\overline{\partial_{\tau_{2n}}\mathcal{O}'(\tau_{2n})}\rangle\,.
\end{empheq}
The NL choice in \eqref{eq: Neumann limit} also suggests considering the operators obtained by taking the source to be the derivative of the Hilbert transform, i.e.  $J''(\tau)\equiv-\partial_{\tau}\hat{J}(\tau)$, in which case one has
\begin{empheq}{alignat=7}
    \delta S&\sim\int_{-\infty}^{\infty}d\tau\langle\bar{\mathcal{O}}(\tau)\rangle\delta J(\tau)&&=\int_{-\infty}^{\infty}d\tau\langle\bar{\mathcal{O}}''(\tau)\rangle\delta J''(\tau)\,,
\end{empheq}
with $\langle\bar{\mathcal{O}}(\tau)\rangle=-\langle\partial_{\tau}\hat{\bar{\mathcal{O}}}''(\tau)\rangle$. An analogous application of the chain rule as in \eqref{chain} gives
\begin{empheq}{alignat=7}
    \langle\mathcal{O}(\tau_1)\cdots\bar{\mathcal{O}}(\tau_{2n})\rangle&=\langle\partial_{\tau_1}\hat{\mathcal{O}}''(\tau_1)\cdots\partial_{\tau_{2n}}\hat{\bar{\mathcal{O}}}''(\tau_{2n})\rangle&&=\langle\partial_{\tau_1}\hat{\mathcal{O}}''(\tau_1)\cdots\overline{\partial_{\tau_{2n}}\hat{\mathcal{O}}''(\tau_{2n})}\rangle\,.
\end{empheq}
These relations are crucial for correctly interpreting the $\chi=\frac{\pi}{2}$ limit of the interpolating boundary conditions: it will yield correlation functions that are related to those in the Neumann case by taking the Hilbert transform and/or (NL/L) a derivative with respect to each of the insertion points. In particular, we expect that
\begin{empheq}{alignat=7}\label{eq: 2-point expected}
	\langle\mathcal{O}(\tau_1)\bar{\mathcal{O}}(\tau_2)\rangle\Big|_{\chi=\frac{\pi}{2}}&=\langle\partial_{\tau_1}\mathcal{O}_N(\tau_1)\overline{\partial_{\tau_2}\mathcal{O}_N(\tau_2)}\rangle\left\{
	\begin{array}{>{\displaystyle}r>{\displaystyle}l}
		1 & \quad\textrm{L}
		\\\\
		-1 & \quad\textrm{NL}
	\end{array}
	\right.\,,
\end{empheq}
where we have used $\bar{\hat{f}}(\tau)=-\hat{\bar{f}}(\tau)$ and that successive Hilbert transforms of $f(\tau_1-\tau_2)$ over $\tau_1$ and $\tau_2$ give back $f(\tau_1-\tau_2)$. We will confirm this expectation below.

\subsection{$2$-point function}\label{subsec: 2-point}
The boundary-to-bulk propagator $K(z,\tau;\tau')$ for a massless scalar field is defined such that the regular solution to the equation of motion is expressed as
\begin{empheq}{alignat=7}\label{eq: definition K}	
	\phi(z,\tau)&=\int_{-\infty}^{\infty}d\tau'K(z,\tau;\tau')J(\tau')\,,
	&\qquad
	\Box K(z,\tau;\tau')&=0\,,
\end{empheq}
where $J(\tau)$ is the boundary data that is fixed in the variational problem. In the case of Dirichlet and Neumann boundary conditions they read 
\begin{empheq}{alignat=7}\label{eq: KD and KN}
	K_D(z,\tau;\tau')&=\frac{1}{\pi}\frac{z}{z^2+(\tau-\tau')^2}\,,
	&\qquad
	K_N(z,\tau;\tau')&=\frac{1}{2\pi}\ln\left(z^2+(\tau-\tau')^2\right)\,,
\end{empheq}
and satisfy
\begin{empheq}{alignat=7}
	K_D(z,\tau;\tau')&\underset{z\to0}{\longrightarrow}\delta(\tau-\tau')\,,
	&\qquad
	\partial_zK_N(z,\tau;\tau')&\underset{z\to0}{\longrightarrow}\delta(\tau-\tau')\,,
\end{empheq}
as appropriate for $J_D(\tau)=\alpha(\tau)$ and $J_N(\tau)=\beta(\tau)$, respectively. For the case at hand, in order to comply with the boundary condition \eqref{eq: definition J free}, the propagator must behave as
\begin{empheq}{alignat=7}
\label{eq: bc for K}
\cos\chi\,K(z,\tau;\tau')+\sin\chi\,\int_{-\infty}^{\infty}d\tau''\partial_zK(z,\tau'';\tau')g(\tau-\tau'')&\underset{z\to0}{\longrightarrow}\delta(\tau-\tau')\,.
\end{empheq}
We readily find
\begin{empheq}{alignat=7}\label{eq: K cases}
	K(z,\tau;\tau')&=\left\{
	\begin{array}{>{\displaystyle}c>{\displaystyle}l}
		\cos\chi\,K_D(z,\tau;\tau')+\sin\chi\,\partial_{\tau}K_N(z,\tau;\tau')& \quad\textrm{L}
		\\\\
		\frac{K_D(z,\tau;\tau')}{\cos\chi+\sin\chi} & \quad\textrm{NL}
	\end{array}
	\right.\,.
\end{empheq}
It is easy to check that 
\begin{empheq}{alignat=7}
	\cos\chi\,K(z,\tau;\tau')+\sin\chi\int_{-\infty}^{\infty}d\tau''\partial_zK(z,\tau'';\tau')g(\tau-\tau'')&=K_D(z,\tau;\tau')\,,
\end{empheq}
thus verifying \eqref{eq: bc for K} for both L and NL boundary conditions.

From \eqref{eq: vev Dirichlet} and \eqref{eq: vev Neumann} we find that the $2$-point functions for Dirichlet and Neumann boundary conditions are given by
\begin{empheq}{alignat=7}\label{DvsN}
	\langle\mathcal{O}_D(\tau_1)\bar{\mathcal{O}}_D(\tau_2)\rangle&=\frac{1}{\pi}\frac{1}{(\tau_1-\tau_2)^2}\,,
	\qquad
	\langle\mathcal{O}_N(\tau_1)\bar{\mathcal{O}}_N(\tau_2)\rangle&=-\frac{1}{\pi}\ln\left(|\tau_2-\tau_1|\right)\,.	
\end{empheq}
In the Neumann case, the dual ($\Delta_{{\cal O}_N}=0$) operator does not transform as a proper primary. As explained above, it is its derivative that is well-behaved, that is,
\begin{empheq}{alignat=7}
 \langle\partial_{\tau_1}\mathcal{O}_N(\tau_1)\overline{\partial_{\tau_2}\mathcal{O}_N(\tau_2)}\rangle&=-\partial_{\tau_1}\partial_{\tau_2}\langle\mathcal{O}_N(\tau_1)\bar{\mathcal{O}}_N(\tau_2)\rangle&&=\frac{1}{\pi}\frac{1}{(\tau_1-\tau_2)^2}\,,
\end{empheq}
where we have used that $\overline{\partial_{\tau}\mathcal{O}_N(\tau)}=-\partial_{\tau}\bar{\mathcal{O}}_N(\tau)$. Similarly, according to \eqref{eq: definition vev free}, the $2$-point functions for the interpolating boundary conditions are
\begin{empheq}{alignat=7}
	\langle\mathcal{O}(\tau_1)\bar{\mathcal{O}}(\tau_2)\rangle&=\frac{1}{\pi}\frac{1}{(\tau_1-\tau_2)^2}\times\left\{
	\begin{array}{>{\displaystyle}c>{\displaystyle}l}
		1 & \qquad\textrm{L}
		\\\\
		\frac{\cos\chi-\sin\chi}{\cos\chi+\sin\chi} & \qquad\textrm{NL}
	\end{array}
	\right.\,.
	\label{sign}
\end{empheq}
As advertised, these expressions correctly reproduce the Dirichlet and Neumann limits. In the version discussed in \cite{Correa_2020}, the L case exhibited a contact term that is now absent from the $2$-point function. This slight difference is explained by the presence of the term $\partial_{\tau'}\bar{J}(\tau')\partial_{\tau}J(\tau)$ in the boundary action \eqref{eq: Sbdry free contact}. The same term is also responsible for yielding a vev that appropriately interpolates between $\beta(\tau)$ (Dirichlet) and $\partial_{\tau}\alpha(\tau)$ (Neumann).

\subsection{$4$-point function}\label{subsec: 4-point}
The computation of $4$-point functions depends on the precise form of the quartic interactions.  In the following we consider the Nambu-Goto action expanded around an open $AdS_2$ worldsheet ending on a straight line at the $AdS_4\times \mathbb{CP}^3$ boundary. The fluctuations along the $\mathbb{CP}^3$ directions are massless, and we focus on a single complex combination of these. To fourth order in $\delta X\sim\lambda^{-\frac{1}{4}}\phi$, the Nambu-Goto action in static gauge becomes \cite{Bianchi_2020}
\begin{empheq}{alignat=7}\label{eq: S quartic}
	S_{\lambda}&=\int d^2x\sqrt{g}\left[|\partial\phi|^2-\sqrt{\frac{2}{\lambda}}\left(|\partial\phi|^2|\phi|^2+\frac{1}{2}|\partial\phi|^4\right)\right]\,,
\end{empheq}
where $g_{\mu\nu}$ is the $AdS_2$ worldsheet metric \eqref{eq: AdS2 metric}, $\lambda=2T^2$ is the ABJM 't Hooft coupling and $T$ is the effective string tension.

A crucial aspect of the action \eqref{eq: S quartic} is that the quartic interactions involve derivatives of the field. This forces us to revisit the variational problem. Indeed, we now have
\begin{empheq}{alignat=7}
	\delta S_{\lambda}&=-\int_{-\infty}^{\infty}d\tau\left(\bar{\eta}(\tau)\delta\alpha(\tau)+\eta(\tau)\delta\bar{\alpha}(\tau)\right)\,,
\end{empheq}
with
\begin{empheq}{alignat=7}\label{eq: definition of eta}
	\eta(\tau)&\equiv\beta(\tau)\left(1-\sqrt{\frac{2}{\lambda}}|\alpha(\tau)|^2\right)\,.
\end{empheq}
As with $S_0$, this action is appropriate for the Dirichlet problem in which $\alpha(\tau)$ is fixed. The natural analogue of Neumann boundary conditions corresponds to fixing $\eta(\tau)$ via
\begin{empheq}{alignat=7}
	S^N_{\textrm{bdry}}&=\int_{-\infty}^{\infty}d\tau\left(\alpha(\tau)\bar{\eta}(\tau)+\bar{\alpha}(\tau)\eta(\tau)\right)\,.
\end{empheq}
In order to impose our proposal of interpolating boundary conditions we add to $S_\lambda$ the term 
\begin{empheq}{alignat=7}\label{eq: Sbdry quartic}
	S_{\textrm{bdry}}&=-\cos\chi\sin\chi\int_{-\infty}^{\infty}d\tau\int_{-\infty}^{\infty}d\tau'\left(\bar{\eta}(\tau)\eta(\tau')-\partial_{\tau'}\bar{\alpha}(\tau')\partial_{\tau}\alpha(\tau)\right)g(\tau-\tau')
	\cr
	&\hspace{11pt}+\sin^2\chi\int_{-\infty}^{\infty}d\tau\left(\bar{\alpha}(\tau)\eta(\tau)+\alpha(\tau)\bar{\eta}(\tau)\right)
	\cr
	&=-\tan\chi\int_{-\infty}^{\infty}d\tau\int_{-\infty}^{\infty}d\tau'\left(\bar{\eta}(\tau)\eta(\tau')-\partial_{\tau'}\bar{J}(\tau')\partial_{\tau}J(\tau)\right)g(\tau-\tau')\,,
\end{empheq}
with $g(\tau)$ as before and
\begin{empheq}{alignat=7}\label{eq: definition J quartic}
	J(\tau)&\equiv\cos\chi\,\alpha(\tau)+\sin\chi\int_{-\infty}^{\infty}d\tau'\eta(\tau')g(\tau-\tau')\,.
\end{empheq}
The $1$-point function then reads
\begin{empheq}{alignat=7}\label{eq: definition vev quartic}
	\langle\bar{\mathcal{O}}(\tau)\rangle&\equiv\cos\chi\,\bar{\eta}(\tau)+\sin\chi\int_{-\infty}^{\infty}d\tau'\partial_{\tau'}\bar{\alpha}(\tau')\partial_{\tau}g(\tau-\tau')
	\cr
	&=\frac{1}{\cos\chi}\left(\bar{\eta}(\tau)+\sin\chi\int_{-\infty}^{\infty}d\tau'\partial_{\tau'}\bar{J}(\tau')\partial_{\tau}g(\tau-\tau')\right)\,.
\end{empheq}
We see that the derivative nature of the quartic potential not only alters the expression for the vev but also requires us to modify the definition of the source (c.f. \eqref{eq: definition J free}, \eqref{eq: definition vev free}, \eqref{eq: Sbdry free}). Fortunately, the modification is quite simple, we just need to replace $\beta(\tau)\to\eta(\tau)$ in every expression. It is important to mention that the boundary condition \eqref{eq: definition J quartic} is now non-linear in the field and depends on the coupling $\lambda^{-\frac{1}{2}}$.

In Appendix \ref{app: derivation} we show that, in spite of the above modifications to the source and vev, the Witten diagram prescription for computing the $4$-point correlation function still works in the usual way: (i) replace each field in the quartic vertex \eqref{eq: S quartic} by the bulk-to-boundary propagator $K(z,\tau;\tau')$ satisfying the appropriate boundary conditions and (ii) symmetrize the insertion points. The expression arising from the connected diagram depicted in figure \ref{fig: Witten diagrams} is given by
\begin{figure}[t!]
\begin{center}
\begin{tikzpicture}
\filldraw [color=black,fill=blue!10] (0,0) circle (1cm);
\draw (-0.707,0.707) -- (0.707,-0.707);
\draw (-0.707,-0.707) -- (0.707,0.707);
\draw [->] (1.25,0) -- (0.25,0);
\filldraw [color=black,fill=white] (-0.707,0.707) circle (2pt);
\filldraw [color=black,fill=white]  (0.707,0.707) circle (2pt);
\filldraw [color=black,fill=white]  (-0.707,-0.707) circle (2pt);
\filldraw [color=black,fill=white]  (0.707,-0.707) circle (2pt);
\filldraw [color=black,fill=black]  (0,0) circle (2pt);
\draw (3.25,0) node {$\sqrt{\frac{2}{\lambda}}\left(|\partial\phi|^2|\phi|^2+\frac{1}{2}|\partial\phi|^4\right)$};
\draw (-1.25,1) node {$\mathcal{O}(\tau_1)$};
\draw (1.25,1) node {$\bar{\mathcal{O}}(\tau_2)$};
\draw (1.25,-1) node {$\mathcal{O}(\tau_3)$};
\draw (-1.25,-1) node {$\bar{\mathcal{O}}(\tau_4)$};
\end{tikzpicture}
\end{center}
\caption{Tree-level diagram coming from the expansion of the Nambu-Goto action.}
\label{fig: Witten diagrams}
\end{figure}
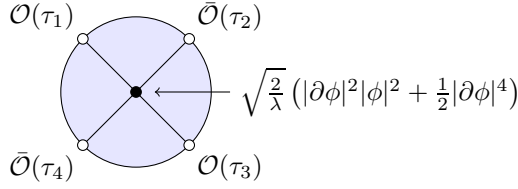
\begin{empheq}{alignat=7}
	\sqrt{\frac{\lambda}{2}}\langle\mathcal{O}(\tau_1)\bar{\mathcal{O}}(\tau_2)\mathcal{O}(\tau_3)\bar{\mathcal{O}}(\tau_4)\rangle&=T(\tau_1,\tau_2,\tau_3,\tau_4)+(\tau_1\leftrightarrow\tau_3)+(\tau_2\leftrightarrow\tau_4)+(\tau_1,\tau_2\leftrightarrow\tau_3,\tau_4)\,,
\end{empheq}
where
\begin{empheq}{alignat=7}
    T(\tau_1,\tau_2,\tau_3,\tau_4)&=U(\tau_1,\tau_2,\tau_3,\tau_4)+V(\tau_1,\tau_2,\tau_3,\tau_4)\,,
\end{empheq}
and
\begin{empheq}{alignat=7}
    \label{eq: integral U}
	U(\tau_1,\tau_2,\tau_3,\tau_4)&=\int\frac{d\tau dz}{z^2}\partial_{\mu}\bar{K}(z,\tau;\tau_1)\partial^{\mu}K(z,\tau;\tau_2)\bar{K}(z,\tau;\tau_3)K(z,\tau;\tau_4)\,,
	\\\label{eq: integral V}
	V(\tau_1,\tau_2,\tau_3,\tau_4)&=\frac{1}{2}\int\frac{d\tau dz}{z^2}\partial_{\mu}\bar{K}(z,\tau;\tau_1)\partial^{\mu}K(z,\tau;\tau_2)\partial_{\nu}\bar{K}(z,\tau;\tau_3)\partial^{\nu}K(z,\tau;\tau_4)\,.
\end{empheq}
These integrals can be computed using the method of residues. In terms of the invariant cross-ratio
\begin{empheq}{alignat=7}
	u&\equiv\frac{\tau_{12}\tau_{34}}{\tau_{13}\tau_{24}}\,,
	&\qquad
	\tau_{ij}&\equiv\tau_i-\tau_j\,,
\end{empheq}
we find
\begin{empheq}{alignat=7}
	\sqrt{\frac{\lambda}{2}}\langle\mathcal{O}(\tau_1)\bar{\mathcal{O}}(\tau_2)\mathcal{O}(\tau_3)\bar{\mathcal{O}}(\tau_4)\rangle_{\textrm{L}}&=\frac{A(u)+\cos(2\chi)B(u)+\cos(4\chi)C(u)+\sin(4\chi)S(\tau_i)}{2\pi^3\tau_{13}^2\tau_{24}^2}
    \cr
	&\hspace{11pt}+\frac{\sin^2\chi\cos(2\chi)}{\pi^3}\left(\frac{1}{\tau_{12}\tau_{14}}-\frac{1}{\tau_{23}\tau_{34}}\right)\left(\frac{1}{\tau_{12}\tau_{23}}-\frac{1}{\tau_{14}\tau_{34}}\right)\,,
\label{4ptL}
\end{empheq}
and
\begin{empheq}{alignat=7}
	\sqrt{\frac{\lambda}{2}}\langle\mathcal{O}(\tau_1)\bar{\mathcal{O}}(\tau_2)\mathcal{O}(\tau_3)\bar{\mathcal{O}}(\tau_4)\rangle_{\textrm{NL}}&=\frac{1}{\left(\cos\chi+\sin\chi\right)^4}\frac{A(u)+B(u)+C(u)}{2\pi^3\tau_{13}^2\tau_{24}^2}\,,
	\label{4ptNL}
\end{empheq}
where
\begin{empheq}{alignat=7}
    A(u)&=-\frac{1}{u^2(1-u)^2}+\frac{u-3}{4(1-u)^3}\ln(u^2)-\frac{2+u}{4u^3}\ln((1-u)^2)\,,
	\cr
	B(u)&=\frac{2}{u(1-u)}\,,\qquad C(u)=-2-\frac{1}{u(1-u)}+\frac{1}{2}\left(2u-1\right)\ln\left(\frac{u^2}{(1-u)^2}\right)\,,
    \cr
    S(\tau_1,\tau_2,\tau_3,\tau_4)&=(1-2u)\Big[\textrm{sign}(\tau_{12})+\textrm{sign}(\tau_{23})+\textrm{sign}(\tau_{34})+\textrm{sign}(\tau_{41})\Big]\frac{\pi}{2}\,.
\end{empheq}
It is possible to write $A(u)$, $B(u)$ and $C(u)$ in terms of conformal invariant $D$-functions \cite{Beccaria:2019dws}. Also, one can check that the function $S(\tau_1,\tau_2,\tau_3,\tau_4)$ is invariant under the full conformal group, including the inversion $\tau\to-1/\tau$. Unlike $u$, it is not invariant under
$\tau\to-\tau$ and $\tau\to1/\tau$ separately. Finally, the terms appearing in the second line of \eqref{4ptL} cannot be written in terms of the cross ratio $u$ and therefore spoil the full conformal covariance of the 4-point function in the L-case. This is the main result of the paper, which we further discuss in the next section. As expected, the $\chi=0$ limit reduces to the Dirichlet case. For $\chi=\frac{\pi}{2}$ our result coincides with that found in \cite{Beccaria:2019dws} for Neumann boundary conditions.

\section{Discussion}\label{sec: discussion}

Motivated by the existence of a one-parameter family of supersymmetric Wilson Loops in ABJM theory,
in this paper we studied two examples of interpolating scale invariant boundary conditions for interacting massless scalar fields in $AdS_2$. 
In order to test if they preserve full conformal symmetry, we computed 4-point correlation functions holographically. Quartic interaction terms were derived from the expansion of the Nambu-Goto action around an open string worldsheet in $AdS_4\times \mathbb{CP}^3$. These terms involve derivatives which induce additional boundary terms in the action. We verified that a first-principles derivation of the 4-point function coincides with the Witten diagrams prescription.

For scalar fields satisfying the L boundary condition, previously discussed in \cite{Correa_2020}, the expression for the 2-point correlation function \eqref{sign} is conformally covariant. However, the holographic 4-point function shown in \eqref{4ptL} cannot be interpreted as that of primary operators in a CFT$_1$. This failure of conformal covariance   is reminiscent of the one  found for the pure Neumann case, studied in \cite{Beccaria:2019dws} in the context of ordinary Wilson Loops in ${\cal N} = 4$ super Yang-Mills. In that reference the anomalous factor disappeared after integrating over the position in $S^5$ around which the Nambu-Goto action was expanded. No such integration appears to be justified in the ABJM case.

On the contrary, in view of the results \eqref{sign} and \eqref{4ptNL},   massless scalars satisfying NL boundary conditions are dual representations of $\Delta = 1$ primaries in the defect CFT$_1$.  The values $\chi=\pm\frac{\pi}{4}$ of the interpolating parameter are special: for $\chi=-\frac{\pi}{4}$ the combination   we  source in \eqref{eq: diff J free} is identically zero, whereas for $\chi=\frac{\pi}{4}$ the vev of the dual operator  in the presence of non-zero sources vanishes. Notice that these two statements, which are a consequence of \eqref{betahatalpha}, are only valid at leading order in $\lambda^{-1/2}$. The fact that the 2-point correlator vanishes for $\chi=\frac{\pi}{4}$ and becomes negative beyond this point seems puzzling, as it would imply a violation of unitarity. We claim, however, that in the range $\frac{\pi}4 <\chi\le \frac{\pi}2$ the LHS of \eqref{eq: definition J free} should be interpreted as the Hilbert transform of the source and not as the source itself. The physical result is then obtained by transforming back with respect to $\tau_1$ and $\tau_2$, a procedure that yields a positive 2-point function. This interpretation becomes evident at the endpoint $\chi=\pi/2$, as argued at the begginning of section \ref{sec: correlation functions}.

The reasons why the 4-point function \eqref{4ptNL} is consistent with conformal covariance but \eqref{4ptL} is not are not clear to us at the moment. One difference between the L and NL boundary conditions, perhaps relevant to understand this issue, is that the former break invariance under parity transformation $\tau \to -\tau$ whereas the latter do not.

In conclusion, out of the two interpolating boundary conditions we have studied, only the NL ones lead to conformally covariant 4-point correlation functions of primary operators. The failure of conformal covariance for the L-boundary conditions case might appear surprising, as the interpolating parameter $\chi$ is dimensionless and the correlation functions are consistent with scale invariance.  
What are the implications of this for the problem of identifying the dual description of the supersymmetric family of interpolating Wilson Loops in ABJM theory  \cite{Ouyang:2015bmy,Ouyang:2015iza}? In principle, this does not necessarily imply that L-boundary condition should be rejected or that the NL-boundary condition should be preferred. The field theory arguments mentioned in the introduction suggest that the interpolating parameter in the family of Wilson loops should be regarded as an exactly marginal deformation\footnote{This interpretation applies when perturbative computations are done at framing $f=1$.  Recently, the same problem was considered with a regularization scheme that uses framing $f=0$ \cite{Castiglioni:2022yes}, in which case the interpolation was interpreted as an RG flow. An interesting open question is what is in correspondence to framing in the dual string theory description.}. However, scale invariance does not imply full conformal symmetry. In order to make a more assertive proposal for the dual representation of the interpolating Wilson loops it would be necessary to understand the conformal transformation properties in the field theory description. For example, it would be interesting to study perturbatively 4-point correlations functions of insertions along those Wilson loops and appraise their conformal covariance in the field theory side. It was shown in \cite{Correa_2020}  that the L-boundary condition describes configurations preserving 4 real supersymmetries, matching the number of supersymmetries of the interpolating Wilson loops. The consistency of  supersymmetry transformations with the NL-boundary condition is another interesting problem that remains to be explored.

Another more speculative possibility is that, instead of being broken,  conformal covariance of the L-boundary condition could be realized in a more intricate way. The anomalous terms in \eqref{4ptL}, although inappropriate for a 4-point correlation function of primary operators of scale dimension $\Delta=1$, would appear in  4-point functions involving operators of scale dimensions $\Delta=3/2$ and $\Delta=1/2$. The appearance of fractional scale dimension for bosonic excitations could point towards derivatives of fractional order, as the ones defined in \cite{Khodaee:2017tbk}. These describe fields in CFT$_1$ that transform under non-local representations of $sl(2)$. Therefore, it could be interesting to explore the possibility of relating our L-boundary condition to excitations in those more generic representations.


Finally, in a general context not related to Wilson loops, one could wonder whether the interpolating boundary conditions presented here can be generalized to scalar fields in $AdS_{d+1}$. Indeed, consider a massive field with $\Delta_+ - \Delta_- =1$; any other masses for which $\Delta_+ - \Delta_-$ equals a larger integer lie outside the BF window. For the L-choice of boundary conditions, the $\tau$-derivative could be generalized to a directional derivative on the $d$-dimensional boundary. This, of course, would break rotational invariance as well as the $d$-dimensional conformal invariance. In order to generalize the NL-boundary condition, we could use the following non-local first order differential operator:\footnote{This definition follows from the Fourier transform of the Laplace operator.}
\begin{equation}
    \sqrt{\nabla^2} f(x) \equiv -\frac{\Gamma(\tfrac{d+1}{2})}{\pi^{\tfrac{d+1}{2}}} \int d^dx' \frac{f(x')}{|x-x'|^{d+1}}\,.
\end{equation}
Since $\sqrt{\partial_\tau^2}f(\tau)=\partial_\tau\hat f(\tau)$ for $d=1$, a natural generalization of the NL boundary condition could be
	\begin{equation}
	    \partial_{\tau}\hat J(\tau) = 	\cos\chi\,\partial_{\tau}\hat \alpha(\tau)
	    -\sin\chi\,\beta(\tau)\qquad
	    \mapsto
	    \qquad
	    \sqrt{\nabla^2} J(x) = \cos\chi\,\sqrt{\nabla^2}\alpha(x)
	    -\sin\chi\,\beta(x)\,.
	\end{equation}
It would be interesting to further analyze these higher-dimensional generalizations and the correlation functions that can be obtained from them.
\section*{Acknowledgments}
We would like to thank  Fernando Alday, Max Ba\~nados, V\'ictor Giraldo Rivera, Mart\'in Lagares and Rodrigo de Le\'on Ard\'on for useful discussions on this problem.  This work was supported by PICT 2020-03749, PICT 2020-03826, PIP 02229,  UNLP X791, UNLP X910 and PUE084 “Búsqueda de nueva física”. DHC and AF would like to acknowledge support from the ICTP through the Associates Programme (2020-2025 and 2022-2027) and GAS would like to acknowledge support from the ICTP visiting programme. The work of AFCG and AF is supported by CONICYT FONDECYT Regular \#1201145 and ANID/ACT210100 Anillo Grant ``Holography and its applications to High Energy Physics, Quantum Gravity and Condensed Matter Systems.''
\appendix

\section{Hilbert Transform}\label{app: Hilbert Transform}
For any function $f(\tau)$ the Hilbert transform $\hat{f}(\tau)$ is defined as \cite{King_2009}
\begin{empheq}{alignat=7}
	\hat{f}(\tau)&\equiv\frac{1}{\pi}\textrm{p.v.}\int_{-\infty}^{\infty}dt\frac{f(t)}{\tau-t}&&=\frac{1}{\pi}\lim_{\epsilon\to0}\int_{-\infty}^{\infty}dt\,f(t)\frac{(\tau-t)}{\epsilon^2+(\tau-t)^2}\,,
\end{empheq}
where p.v. stands for Cauchy principal value. This can also be written as
\begin{empheq}{alignat=7}
	\hat{f}(\tau)&=\frac{d}{d\tau}\left(\frac{1}{\pi}\textrm{p.v.}\int_{-\infty}^{\infty}dt\,f(t)\ln\left(|\tau-t|\right)\right)&&=\frac{d}{d\tau}\left(\frac{1}{2\pi}\lim_{\epsilon\to0}\int_{-\infty}^{\infty}dt\,f(t)\ln\left(\epsilon^2+(\tau-t)^2\right)\right)\,.
\end{empheq}
These definitions hold provided that the integrals exists. More generally, one can consider $f(\tau)$ to be a distribution. In Fourier space the Hilbert transform is a multiplicative operator, namely,
\begin{empheq}{alignat=7}
	\mathcal{F}(\hat{f}\hspace{1pt})(w)&=-i\,\textrm{sign}(w)\mathcal{F}(f\hspace{0.5pt})(w)\,.
\end{empheq}
A few transform pairs are listed in table \ref{Table: HT pairs} and some useful properties are shown in table \ref{Table: HT properties}.
\begin{table}[h]
\begin{displaymath}
\begin{array}{*2{|>{\displaystyle}c}|}
	\hline
	\vphantom{\Big|}
	\textrm{Function} & \textrm{Hilbert transform}
	\\
	\hline\hline
	\vphantom{\Bigg|}
	\textrm{constant} & 0
	\\
	\hline
	\vphantom{\Bigg|}
	\delta(\tau) &\frac{1}{\pi\tau}
	\\
	\hline
	\vphantom{\Bigg|}
	\ln(|\tau|) & -\frac{\pi}{2}\textrm{sign}(\tau)
	\\
	\hline
	\vphantom{\Bigg|}
	e^{iw\tau} & -i\,\textrm{sign}(w)e^{iw\tau}
	\\
	\hline
	\vphantom{\Bigg|}
	\frac{1}{2}\ln\left(z^2+\tau^2\right) & -\tan^{-1}\left(\frac{\tau}{|z|}\right)
	\\
	\hline
	\vphantom{\Bigg|}
	\frac{z}{z^2+\tau^2} & \textrm{sign}(z)\frac{\tau}{z^2+\tau^2}
	\\
	\hline
\end{array}
\end{displaymath}
\caption{Examples of Hilbert transform pairs.}
\label{Table: HT pairs}
\end{table}
\newpage
\begin{table}[h]
\begin{displaymath}
\begin{array}{*3{|>{\displaystyle}c}|}
	\hline
	\multicolumn{2}{|c|}{\textrm{Property}} & \vphantom{\Big|}\textrm{Comments}
	\\
	\hline\hline
	\multicolumn{2}{|>{\displaystyle}c|}{\int_{-\infty}^{\infty}d\tau\,\hat{f}(\tau)g(\tau)=-\int_{-\infty}^{\infty}d\tau\,f(\tau)\hat{g}(\tau)} & \vphantom{\Bigg|}\textrm{anti-self adjoint}
	\\
	\hline\hline
	\vphantom{\Big|}
	\textrm{Function} & \textrm{Hilbert transform} & 
	\\
	\hline\hline
	\vphantom{\Bigg|}
	\hat{f}(\tau) & -f(\tau) & \textrm{inverse}
	\\
	\hline
	\vphantom{\Bigg|}
	\partial_{\tau}f(\tau) & \partial_{\tau}\hat{f}(\tau) & \textrm{derivative}
	\\
	\hline
	\vphantom{\Bigg|}
	(f*g)(\tau) & (\hat{f}*g)(\tau)=(f*\hat{g})(\tau) & \textrm{convolution}
	\\
	\hline
	\vphantom{\Bigg|}
	\bar{f}(\tau) & -\bar{\hat{f}}(\tau) & \bar{f}(\tau)\equiv f(-\tau)^*
	\\
	\hline
	\vphantom{\Bigg|}
	f(\tau-a) & \hat{f}(\tau-a) & \textrm{translations}
	\\
	\hline
	\vphantom{\Bigg|}
	f(a\tau) & \textrm{sign}(a)\hat{f}(a\tau) &\textrm{rescalings/parity}
	\\
	\hline
	\vphantom{\Bigg|}
	f(-1/\tau) & \hat{f}(-1/\tau)-\hat{f}(0) &\textrm{inversion}
	\\
	\hline
\end{array}
\end{displaymath}
\caption{Useful Hilbert transform properties.}
\label{Table: HT properties}
\end{table}

\section{Bulk-to-bulk propagators}\label{app: G}
In AdS/CFT the computation of tree-level correlation functions involves solving non-linear equations of motion, which in turn requires knowing both the bulk-to-boundary and bulk-to-bulk propagators. For a massless scalar field the latter is defined by
\begin{empheq}{alignat=7}
	\Box_xG(x;x')&=\frac{1}{\sqrt{g}}\delta(x;x')\,.
\end{empheq}
For the $AdS_2$ metric in conformal gauge \eqref{eq: AdS2 metric}, the Dirichlet and Neumann propagators  coincide with those in flat space,
\begin{empheq}{alignat=7}
	G_{D/N}(z,\tau;z',\tau')&=\frac{1}{4\pi}\left(\ln\left((z-z')^2+(\tau-\tau')^2\right)\mp\ln\left((z+z')^2+(\tau-\tau')^2\right)\right)\,,
\end{empheq}
and satisfy
\begin{empheq}{alignat=7}
	G_D(z,\tau;z',\tau')\Big|_{z=0}&=0\,,
	&\qquad
	\partial_{z}G_N(z,\tau;z',\tau')\Big|_{z=0}&=0\,.
\end{empheq}
Moreover, they are related to the  corresponding $AdS_2$ bulk-to-boundary propagators \eqref{eq: KD and KN} by
\begin{empheq}{alignat=7}
	K_D(z',\tau';\tau)&=-\partial_{z}G_D(z,\tau;z',\tau')\Big|_{z=0}\,,
	&\qquad
	K_N(z',\tau';\tau)&=G_N(z,\tau;z',\tau')\Big|_{z=0}\,.
\end{empheq}
This follows directly from Green's third identity
\begin{empheq}{alignat=7}\label{eq: Green}
	\phi(z',\tau')&=-\int_{-\infty}^{\infty}d\tau\left(\phi(z,\tau)\partial_{z}G(z,\tau;z',\tau')-\partial_{z}\phi(z,\tau)G(z,\tau;z',\tau')\right)\Big|_{z=0}\,,
\end{empheq}
valid for any harmonic function $\phi$ on the upper half-plane. 

In order to identify the correct boundary conditions for $G(x;x')$ in the interpolating case we re-write Green's identity as
\begin{empheq}{alignat=7}\label{eq: Green 2}
	\phi(z',\tau')&=-\int_{-\infty}^{\infty}d\tau\left[\left(\phi(z,\tau)+\tan\chi\int_{-\infty}^{\infty}d\tau''\partial_{z}\phi(z,\tau'')g(\tau-\tau'')\right)\partial_{z}G(z,\tau;z',\tau')\right.
	\cr
	&\hspace{11pt}\left.-\partial_{z}\phi(z,\tau)\left(G(z,\tau;z',\tau')+\tan\chi\int_{-\infty}^{\infty}d\tau''\partial_{z}G(z,\tau'';z',\tau')g(-\tau+\tau'')\right)\right]\Bigg|_{z=0}\,,
\end{empheq}
where we have swapped the order of integration in the last term. Recalling that
\begin{empheq}{alignat=7}
	\Box\phi&=0
	&\qquad\Rightarrow\qquad
	\int_{-\infty}^{\infty}d\tau\,\partial_{z}\phi(z,\tau)\Big|_{z=0}&=0\,,
	\label{intebetacondition}
\end{empheq}
we see that a sufficient condition for the second line in \eqref{eq: Green 2} to vanish is
\begin{empheq}{alignat=7}\label{eq: bc for G}
	\partial_{\tau}\left(\cos\chi\,G(z,\tau;z',\tau')+\sin\chi\int_{-\infty}^{\infty}d\tau''\partial_{z}G(z,\tau'';z',\tau')g(-\tau+\tau'')\right)\Bigg|_{z=0}&=0\,.
\end{empheq}
The field can then be reconstructed from the boundary data using the bulk-to-boundary propagator
\begin{empheq}{alignat=7}\label{eq: relation G and K}
	K(z',\tau';\tau)&=-\frac{1}{\cos\chi}\partial_{z}G(z,\tau;z',\tau')\Big|_{z=0}\,.
\end{empheq}

The bulk-to-bulk Green's functions with  boundary conditions \eqref{eq: bc for G} can be found with the help of the relations
\begin{empheq}{alignat=7}
	\partial_{\tau'}G_{D/N}(z,\tau;z',\tau')&=\partial_{z'}\hat{G}_{N/D}(z,\tau;z',\tau')\,,
	\hspace{15pt}
	\partial_{z'}G_{D/N}(z,\tau;z',\tau')&=-\partial_{\tau'}\hat{G}_{N/D}(z,\tau;z',\tau')\,,
\end{empheq}
valid for $z>z'$, where the Hilbert transforms of the Dirichlet and Neumann propagators read
\begin{empheq}{alignat=7}
	\hat{G}_{D/N}(z,\tau;z',\tau')&=-\frac{1}{2\pi}\left(\tan^{-1}\left(\frac{\tau-\tau'}{|z-z'|}\right)\mp\tan^{-1}\left(\frac{\tau-\tau'}{z+z'}\right)\right)\,.
\end{empheq}
Notice that
\begin{empheq}{alignat=7}
	\hat{G}_D(z,\tau;z',\tau')-\hat{G}_N(z,\tau;z',\tau')&=\frac{1}{\pi}\tan^{-1}\left(\frac{\tau-\tau'}{z+z'}\right)
\end{empheq}
is a harmonic function but
\begin{empheq}{alignat=7}
	\hat{G}_D(z,\tau;z',\tau')+\hat{G}_N(z,\tau;z',\tau')&=-\frac{1}{\pi}\tan^{-1}\left(\frac{\tau-\tau'}{|z-z'|}\right)
\end{empheq}
is not.  We find that
\begin{empheq}{alignat=7}\label{eq: G cases}
	G(x;x')&=\left\{
	\begin{array}{>{\displaystyle}c>{\displaystyle}l}
		c^2_\chi\,G_D(x;x')+s^2_\chi\,G_N(x;x')-c_\chi s_\chi\big(\hat{G}_D(x;x')-\hat{G}_N(x;x')\big) & \quad\textrm{L}
		\\\\
		\frac{c_\chi\,G_D(x;x')+s_\chi\,G_N(x;x')}{c_\chi+s_\chi} & \quad\textrm{NL}
	\end{array}
	\right.\,,
\end{empheq}
where we have abbreviated $c_\chi=\cos\chi$ and $s_\chi=\sin\chi$. Satisfyingly, we find from \eqref{eq: relation G and K} that \eqref{eq: G cases} correctly reproduces \eqref{eq: K cases}.

\section{First Principles Derivation of Witten Diagrams}\label{app: derivation}

According to the AdS/CFT dictionary, at tree-level, the $4$-point function in the dual field theory is given by
\begin{empheq}{alignat=7}
	\langle\mathcal{O}(\tau_1)\bar{\mathcal{O}}(\tau_2)\mathcal{O}(\tau_3)\bar{\mathcal{O}}(\tau_4)\rangle&=-\frac{\delta^4S[J]}{\delta\bar{J}(\tau_1)\delta J(\tau_2)\delta\bar{J}(\tau_3)\delta J(\tau_4)}&&=\frac{1}{\cos\chi}\frac{\delta^3\bar{\eta}(\tau_4)}{\delta\bar{J}(\tau_1)\delta J(\tau_2)\delta\bar{J}(\tau_3)}\,,
\end{empheq}
where we have used \eqref{eq: definition vev quartic} to identify the first derivative of the on-shell action. Notice that the contact term that renders the $2$-point function conformally covariant is linear in $J(\tau)$ and therefore does not affect higher order correlators. 

To find the vev as a function of the source we need to solve the non-linear equation of motion derived from the quartic action \eqref{eq: S quartic} with interpolating boundary conditions \eqref{eq: definition J quartic}.
We do this perturbatively in the coupling $\lambda^{-\frac{1}{2}}$ by expanding the field as
\begin{empheq}{alignat=7}
    \label{expansionphi}
	\phi&=\phi_0+\sqrt{\frac{2}{\lambda}}\phi_1+\mathcal{O}(\lambda^{-1})\,.
\end{empheq}
We find that
\begin{empheq}{alignat=7}\label{eq: perturbative eom}
	\Box\phi_0&=0\,,
	&\qquad
	\Box\phi_1&=J_{\textrm{bulk}}\,,
\end{empheq}
where
\begin{empheq}{alignat=7}
	J_{\textrm{bulk}}&=-\phi_0|\partial\phi_0|^2+\nabla_{\mu}\left(\partial^{\mu}\phi_0|\phi_0|^2\right)+\nabla_{\mu}\left(\partial^{\mu}\phi_0|\partial\phi_0|^2\right)&&=\bar{\phi}_0(\partial\phi_0)^2+\nabla_{\mu}\left(\partial^{\mu}\phi_0|\partial\phi_0|^2\right)\,.
\end{empheq}
The asymptotic coefficients $\alpha(\tau)$ and $\beta(\tau)$ as well as the function $\eta(\tau)$ defined in \eqref{eq: definition of eta} are also expanded as
\begin{empheq}{alignat=7}
    \label{eq: expansion of coeffs}
    \alpha&=\alpha_0+\sqrt{\frac{2}{\lambda}}\alpha_1+\mathcal{O}(\lambda^{-1})\,,
    &\qquad
    \beta&=\beta_0+\sqrt{\frac{2}{\lambda}}\beta_1+\mathcal{O}(\lambda^{-1})\,,
    &\qquad
    \eta&=\eta_0+\sqrt{\frac{2}{\lambda}}\eta_1+\mathcal{O}(\lambda^{-1})\,,
\end{empheq}
with
\begin{empheq}{alignat=7}
	\eta_0&=\beta_0\,,
	&\qquad
	\eta_1&=\beta_1-\beta_0\bar{\alpha}_0\alpha_0\,.
\end{empheq}
Substituting \eqref{eq: expansion of coeffs} into \eqref{eq: definition J quartic} we get
\begin{empheq}{alignat=7}\label{eq: bc phi0}
	\cos\chi\,\alpha_0(\tau)+\sin\chi\int_{-\infty}^{\infty}d\tau'\beta_0(\tau')g(\tau-\tau')&=J(\tau)\,,
	\\\label{eq: bc phi1}
	\cos\chi\,\alpha_1(\tau)+\sin\chi\int_{-\infty}^{\infty}d\tau'\beta_1(\tau')g(\tau-\tau')&=F(\tau)\,,
\end{empheq}
where
\begin{empheq}{alignat=7}
	F(\tau)&=\sin\chi\int_{-\infty}^{\infty}d\tau'\beta_0(\tau')\bar{\alpha}_0(\tau')\alpha_0(\tau')g(\tau-\tau')\,.
\end{empheq}
As usual, the source $J(\tau)$ enters linearly in the boundary condition for $\phi_0$. The novel feature, brought in by the derivative nature of the vertices, is the non-homogeneous boundary condition for the fluctuation $\phi_1$. The classical solution satisfying \eqref{eq: bc phi0} and \eqref{eq: bc phi1} is then given by
\begin{empheq}{alignat=7}
    \label{eq: phi0 in terms of J}
	\phi_0(z,\tau)&=\int_{-\infty}^{\infty}d\tau'K(z,\tau;\tau')J(\tau')\,,
	\\
	\phi_1(z,\tau)&=\int\frac{dz'd\tau'}{z'^2}\bar{G}(z,\tau;z',\tau')J_{\textrm{bulk}}(z',\tau')+\int_{-\infty}^{\infty}d\tau'K(z,\tau;\tau')F(\tau')\,.
\end{empheq}
Here $K(z,\tau;\tau')$ and $G(z,\tau;z',\tau')$ are the bulk-to-boundary and bulk-to-bulk propagators\footnote{The conjugation in $G(z,\tau;z',\tau')$ is due to the interchange $\tau\leftrightarrow\tau'$. A similar 
property holds for $K(z,\tau;\tau')$. } introduced in section \ref{subsec: 2-point} and appendix \ref{app: G}. Notice that $\phi_0$ and $\phi_1$ have the correct asymptotics as a consequence of \eqref{eq: bc for K} and \eqref{eq: bc for G}. 

The term $\eta_0=\beta_0$ in \eqref{eq: expansion of coeffs} does not contribute to the $4$-point function since it is linear in $J$. To compute $\eta_1$ we need
\begin{empheq}{alignat=7}
	\beta_1(\tau)&=\int\frac{dz'd\tau'}{z'^2}\partial_z\bar{G}(z,\tau;z',\tau')\Big|_{z=\epsilon}J_{\textrm{bulk}}(z',\tau')+\int_{-\infty}^{\infty}d\tau'\partial_z K(z,\tau;\tau')\Big|_{z=\epsilon}F(\tau')\,.
\end{empheq}
The first term requires no further manipulation once we recall the relation \eqref{eq: relation G and K} between the bulk-to-bulk and bulk-to-boundary propagators. The second term can be simplified by replacing the definition of $F(\tau)$, swapping the order of integration, and using the boundary condition for $K(z,\tau;\tau')$. In this process, the $\delta$-function appearing in \eqref{eq: bc for K} will cancel against the term $\beta_0\bar{\alpha}_0\alpha_0$ in $\eta_1$. Putting these ingredients together yields
\begin{empheq}{alignat=7}\label{eq: eta1}
	\frac{\eta_1(\tau)}{\cos\chi}&=-\int\frac{dz'd\tau'}{z'^2}J_{\textrm{bulk}}(z',\tau')\bar{K}(z',\tau';\tau)-\int_{-\infty}^{\infty}d\tau'\beta_0(\tau')\bar{\alpha}_0(\tau')\alpha_0(\tau')K(z,\tau;\tau')\Big|_{z=\epsilon}\,.
\end{empheq}
Finally, using the expression \eqref{eq: phi0 in terms of J} for $\phi_0$ we arrive at
\begin{empheq}{alignat=7}
	\frac{\bar{\eta}_1(\tau_4)}{\cos\chi}&=-\int_{-\infty}^{\infty}d\tau_1d\tau_2d\tau_3\,\bar{J}(\tau_1)J(\tau_2)\bar{J}(\tau_3)T'(\tau_1,\tau_2,\tau_3,\tau_4)\,,
\end{empheq}
where
\begin{empheq}{alignat=7}
    T'(\tau_1,\tau_2,\tau_3,\tau_4)&=U'(\tau_1,\tau_2,\tau_3,\tau_4)+V'(\tau_1,\tau_2,\tau_3,\tau_4)+W'(\tau_1,\tau_2,\tau_3,\tau_4)\,,
\end{empheq}
and
\begin{empheq}{alignat=7}
    \label{eq: integral U'}
	U'(\tau_1,\tau_2,\tau_3,\tau_4)&=\int\frac{d\tau dz}{z^2}\partial_{\mu}\bar{K}(z,\tau;\tau_1)K(z,\tau;\tau_2)\partial^{\mu}\bar{K}(z,\tau;\tau_3)K(z,\tau;\tau_4)\,,
	\\\label{eq: integral V'}
	V'(\tau_1,\tau_2,\tau_3,\tau_4)&=\int\frac{d\tau dz}{z^2}\nabla_{\mu}\left(\partial_{\nu}\bar{K}(z,\tau;\tau_1)\partial^{\nu}K(z,\tau;\tau_2)\partial^{\mu}\bar{K}(z,\tau;\tau_3)\right)K(z,\tau;\tau_4)\,,
	\\\label{eq: integral W'}
	W'(\tau_1,\tau_2,\tau_3,\tau_4)&=\int_{-\infty}^{\infty}d\tau\,\partial_z\bar{K}(z,\tau;\tau_1)K(z,\tau;\tau_2)\bar{K}(z,\tau;\tau_3)K(z,\tau;\tau_4)\Big|_{z=\epsilon}\,.
\end{empheq}
The integrals $U'$ and $V'$ clearly originate from the bulk current $J_{\textrm{bulk}}$, while $W'$ arises from the boundary term in \eqref{eq: eta1}. With this the $4$-point function becomes
\begin{empheq}{alignat=7}
	\langle\mathcal{O}(\tau_1)\bar{\mathcal{O}}(\tau_2)\mathcal{O}(\tau_3)\bar{\mathcal{O}}(\tau_4)\rangle&=-\sqrt{\frac{2}{\lambda}}\Big[T'(\tau_1,\tau_2,\tau_3,\tau_4)+(\tau_1\leftrightarrow\tau_3)\Big]\,.
	\label{4ptapp}
\end{empheq}

The equivalence between the above derivation and the standard calculation using Witten diagrams can be shown by a simple manipulation of the integrals \eqref{eq: integral U'}-\eqref{eq: integral W'}. Indeed, one verifies by integration by parts that
\begin{empheq}{alignat=7}
	U'(\tau_1,\tau_2,\tau_3,\tau_4)&=-U(\tau_1,\tau_2,\tau_3,\tau_4)-U(\tau_1,\tau_4,\tau_3,\tau_2)-W'(\tau_1,\tau_2,\tau_3,\tau_4)\,,
	\\
	V'(\tau_1,\tau_2,\tau_3,\tau_4)&=-2V(\tau_1,\tau_2,\tau_3,\tau_4)\,,
\end{empheq}
where $U(\tau_1,\tau_2,\tau_3,\tau_4)$ and $V(\tau_1,\tau_2,\tau_3,\tau_4)$ are defined in \eqref{eq: integral U} and \eqref{eq: integral V}, respectively. Then, using the manifest symmetry $V(\tau_1,\tau_2,\tau_3,\tau_4)=V(\tau_3,\tau_4,\tau_1,\tau_2)$, the $4$-point correlator becomes
\begin{empheq}{alignat=7}
	\langle\mathcal{O}(\tau_1)\bar{\mathcal{O}}(\tau_2)\mathcal{O}(\tau_3)\bar{\mathcal{O}}(\tau_4)\rangle&=\sqrt{\frac{2}{\lambda}}\Big[T(\tau_1,\tau_2,\tau_3,\tau_4)+(\tau_1\leftrightarrow\tau_3)+(\tau_2\leftrightarrow\tau_4)+(\tau_1,\tau_2\leftrightarrow\tau_3,\tau_4)\Big]\,,
\end{empheq}
with
\begin{empheq}{alignat=7}
	T(\tau_1,\tau_2,\tau_3,\tau_4)&=U(\tau_1,\tau_2,\tau_3,\tau_4)+V(\tau_1,\tau_2,\tau_3,\tau_4)\,.
\end{empheq}
This is precisely the expression corresponding to the Witten diagram in figure \ref{fig: Witten diagrams}.
\bibliographystyle{JHEP}
\bibliography{Bib}
\end{document}